\documentclass[10pt]{article}

\usepackage{amsmath}
\usepackage{amssymb}

\usepackage{graphicx}

\usepackage{cite}
\usepackage{color}

\usepackage[labelfont=bf,labelsep=period,justification=raggedright]{caption}

\makeatletter
\renewcommand{\@biblabel}[1]{\quad#1.}
\makeatother

\date{}

\newcommand{\be}{\begin{equation}}
\newcommand{\ee}{\end{equation}}
\newcommand{\ce}{{\it C. elegans\ }}
\newcommand{\cep}{{\it C. elegans}}

\begin{document}

\begin{flushleft}
{\Large \textbf{Colored motifs reveal computational building blocks in the C. elegans brain}}\\\mbox{}\\
Jifeng Qian$^{1}$, Arend Hintze$^{1,2}$, Christoph Adami$^{1,3,4,5\ast}$\\
\mbox{}\\

\bf{1}Keck Graduate Institute, Claremont, CA 91711\\
\bf{2}Computer Science and Engineering, Michigan State University, East Lansing, MI 48824\\
\bf{3}Computation and Neural Systems 139-70, California Institute of Technology, Pasadena, CA 91125\\
\bf{4}Microbiology and Molecular Genetics, Michigan State University, East Lansing, MI 48824\\
\bf{5}BEACON Center for Evolution in Action, Michigan State University, East Lansing, MI 48824\\
$\ast$ E-mail: adami@kgi.edu
\end{flushleft}

\section*{Abstract}
\textbf{Background} 
Complex networks can often be decomposed into less complex sub-networks whose structures can give hints about the functional organization of the network as a whole. However, these structural motifs can only tell one part of the functional story because in this analysis each node and edge is treated on an equal footing. In real networks,  two motifs that are topologically identical but whose nodes perform very different functions will play very different roles in the network. 

\noindent\textbf{Methodology/Principal Findings} Here, we combine structural information derived from the topology of the neuronal network of the nematode {\it C. elegans} with information about the biological function of these nodes, thus {\em coloring} nodes by function. We discover that particular colorations of motifs are significantly more abundant in the worm brain than expected by chance, and have particular computational functions that emphasize the feed-forward structure of information processing in the network, while evading feedback loops. Interneurons are strongly over-represented among the common motifs, supporting the notion that these motifs process and transduce the information from the sensor neurons towards the muscles. Some of the most common motifs identified in the search for significant colored motifs play a crucial role in the system of neurons controlling the worm's locomotion. 

\noindent\textbf{Conclusions/Significance}
The analysis of complex networks in terms of colored motifs combines two independent data sets to generate insight about these networks that cannot be obtained with either data set alone. The method is general and should allow a decomposition of any complex networks into its functional (rather than topological) motifs as long as both wiring and functional information is available.


\section*{Introduction}

Over the last decades, systems biology and network theory have contributed tremendously to our understanding of complex systems~\cite{Barabasi2002,BarabasiOltvai2004,Newmanetal2006,Alon2007}, revealing for example that the topological architecture of the molecular interaction networks within a cell is shared to a large degree by other complex systems, such as the Internet, computer chips and society~\cite{BarabasiOltvai2004}. This insight led to the development of various quantitative tools in network theory to analyze the complex structures within biological networks.

Complex networks like electronic circuits are frequently represented in terms of modules such as operational amplifiers, logical gates and memory, and it is often suggested that biological networks can similarly decomposed into functional modules that have stereotypical functions~\cite{Hartwelletal1999,CallebautRasskin2005}. Because the detection and identification of modules is a notoriously difficult task~\cite{Newman2003}, a different approach focuses on the identification of conserved network motifs~\cite{ShenOrretal2002,Miloetal2002,Riceetal2005}, that is, sub-networks of small size (typically two to five nodes) that are significantly more abundant in a network compared to a control network that had its edges randomly rearranged. The idea behind looking for significant motifs is evolutionary in nature: those motifs that are conducive to the function of the organism within its environment will be preferentially maintained over motifs that are either neutral in function or even detrimental. For example, an analysis of structural motif abundances in a variety of biological networks shows that these abundances can in part be explained by the motifs' robustness to small perturbations~\cite{Prilletal2005}. This thinking equally applies to technological systems that do not evolve according to strict Darwinian rules. For example, a comparison of motif abundances in biological, technological, social, and even word-adjacency networks~\cite{Miloetal2004} shows that these networks can be grouped into clusters that share similar motif abundance profiles.

We analyze motifs in the network of synaptic and gap-junction connections of the neuronal network of the nematode \cep. This network controls one of the most well-understood complex biological systems to date, and  most of the network architecture of the 302 neurons of the hermaphrodite worm is known from experimental work~\cite{Whiteetal1986,HallRussell1991} as well as recent reconstructions~\cite{Varshneyetal2009}. The most up-to-date wiring information covers 279 neurons of the somatic nervous system, excluding 20 neurons of the pharyngeal system and three neurons that appear to be unconnected from the rest~\cite{Varshneyetal2009}.
There are 3,606 edges between these nodes, of which some (the synaptic connections) are directed, while gap-junctions are undirected.   

An analysis of topological motifs in this network has revealed that two major building blocks are significantly overrepresented in the \ce neuronal network: the feedforward loop, and the bi-fan motif~\cite{Reigletal2004}. It is believed that these motifs perform stereotypic functions and play a crucial role in the nematode's descision-making and control~\cite{SpornsKoetter2004,Songetal2005,Varshneyetal2009}. 
However, while there is support for the hypothesis that over-represented motifs point to biological function from the evolutionary conservation of motifs in the yeast protein-protein interaction network~\cite{Wuchtyetal2003}, these conclusions have also been questioned~\cite{Ingrametal2006} on the grounds that topology alone does not contain enough information to predict the function or process, or how biochemical reactions are likely to proceed in biological systems~\cite{Miloetal2002}. 
\begin{figure}[!ht] 
   \centering
   \includegraphics[width=1.5in]{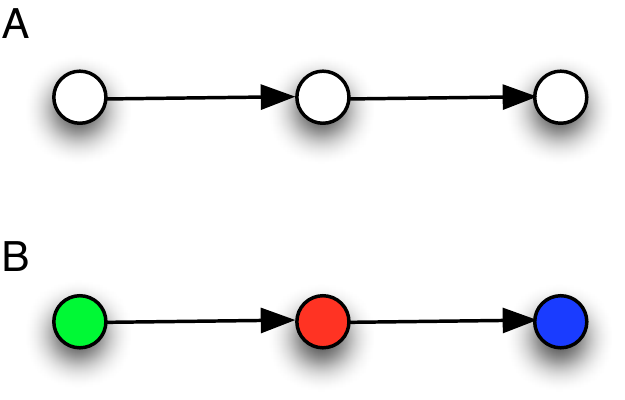} 
   \caption{{\bf Significance of uncolored vs. colored motifs.} (A): Non-significant motif from an uncolored analysis~\protect\cite{Reigletal2004} becomes highly significant (B) if colors are used to attach functional tags to the nodes. Green: sensor neurons, red: interneurons, blue: motor neurons. \label{fig:colmotif}
}
\end{figure}
Indeed, the identification of the feedforward and bifan motifs does not allow us to determine {\em how} these motifs are used, or how they contribute to the worm's behavior. A simple example can illustrate this point: in Fig.~\ref{fig:colmotif}A, we show a three-node motif that was not found to be significantly overrepresented in previous analyses~\cite{SpornsKoetter2004,Reigletal2004,Songetal2005}. However, if we color each neuron according to three possible functional tags such as motorneuron (blue), sensor neuron (green), or interneuron (red), several colored motifs stand out with high significance (see below), among which the motif shown in Fig.~\ref{fig:colmotif}B. The functional significance of this motif is immediately obvious: it relays sensory information via an interneuron towards a muscle.  Indeed, previous studies have shown that the connections between neurons of the three types chosen here are heavily biased: neurons do not connect indiscriminately between types~\cite{Lichtmanetal2008,White2002,Songetal2005,YoshimuraCallaway2005}. Also, an analysis of colored motifs using GO annotations in the yeast protein-protein interaction network~\cite{Leeetal2006} suggests that differently colored motifs are differentially evolutionarily conserved, pointing to a diversity of functional roles for motifs with the same structure.

Here we combine two important data sets for a systematic analysis of the topological {\em and} functional motifs in the \ce brain:  the connection graph and the functional characterization of each neuron~\cite{Varshneyetal2009}. Using these datasets,  the entire \ce neuronal network becomes a {\em colored graph} where nodes represent neurons, edges are connections between neurons, and the color of the node tags the cell-type of the node. It is clear that the choice of the cell-type set (the colors) is crucial for the success of this method, and different choices will produce different results. 
At the same time, the classification into the three cell types is in itself ambiguous, because there are differences of annotation in the literature, and some cells are sometimes annotated as belonging to two classes. Other classifications exist (such as into ten different morphological classes~\cite{AchacosoYamamoto1992}), but the motif analysis of graphs with more than three colors quickly becomes computationally cumbersome. Here, we study the abundance distribution of colored directed motifs of sizes two to four nodes. The number of possible motifs in a network strongly depends on the size of the motif, whether edges are directed, and the number of colors used to tag the nodes (see Table 1).

While an identification of functional motifs can help us understand how the worm uses its neuronal network for signal transduction, we should keep in mind that the worm also uses extrasynaptic signaling for behavior~\cite{ChaseKoelle2007}. Furthermore, several different molecules can modulate synaptic function at a single neuron~\cite{Richmond2007}. Thus, some of the computation that translates signals into actions takes place outside of the connection graph proper, and cannot be explored via a motif analysis. 

\begin{table}[h]
\caption{\bf{Comparison of number of possible colored motifs in {\it C. elegans} neuronal network}}.    %
  \begin{tabular}{@{} l  ccc @{}} 
      & size 2 & size 3 & size 4 \\
   \hline
         UM(1)     & 1/1 & 2/2& 6/6 \\
      DM(1)          & 2/2   &  13/13 & 199/199 \\
      UM(2)        & 3/3  & 10/10&50/50 \\
      UM(3)   & 6/6 & 28/28&201/201 \\
      DM(2) & 7/7   &  86/86& 2,818/2,818 \\	
      DM(3) & 15/15 & 262/273 & 8,310/13,770
   \end{tabular}
   \begin{flushleft}The table shows the {\em actual} numbers of colored motifs of a given size (and directedness) found in {\it C. elegans}Õ neuronal network as well as the theoretically {\em possible} number of colored motifs as a pair of numbers (actual/possible). UM(1): undirected motifs, uni-colored, DM(1): directed motifs, uni-colored, UM(2): undirected motifs, two colors, DM(2): directed motifs, two colors, and so on. 
\end{flushleft}
     \label{tab:Tab1}
 \end{table}

\section*{Results}

\subsection*{Adaptive significance of colored motif distribution}

If the {\em coloration} of a motif (that is, the identity of colors at different positions of the motif) has adaptive significance, we should see a bias in the colored motif distribution with respect to control networks whose color assignments have been scrambled. We extract colored motif abundances from the colored networks by counting all distinct color combinations for each of the structural motifs of size 2, 3, and 4. Of the 279 neurons, 86 are classified as sensor neurons (and colored green in the following), 80 are classified as interneurons (red), and the remainder of 114 neurons are classified as motorneurons (blue)~\cite{Varshneyetal2009}. We stress again that the classification of some neurons is uncertain because other groups~\cite{Wormatlas2002-2010} have classified some neurons as belonging to two types simultaneously, and some neurons' classification is tentative. However, the results presented here do not vary significantly if a few neurons are misclassified.  

\begin{figure}[!ht]
   \centering
  \includegraphics[width=4in]{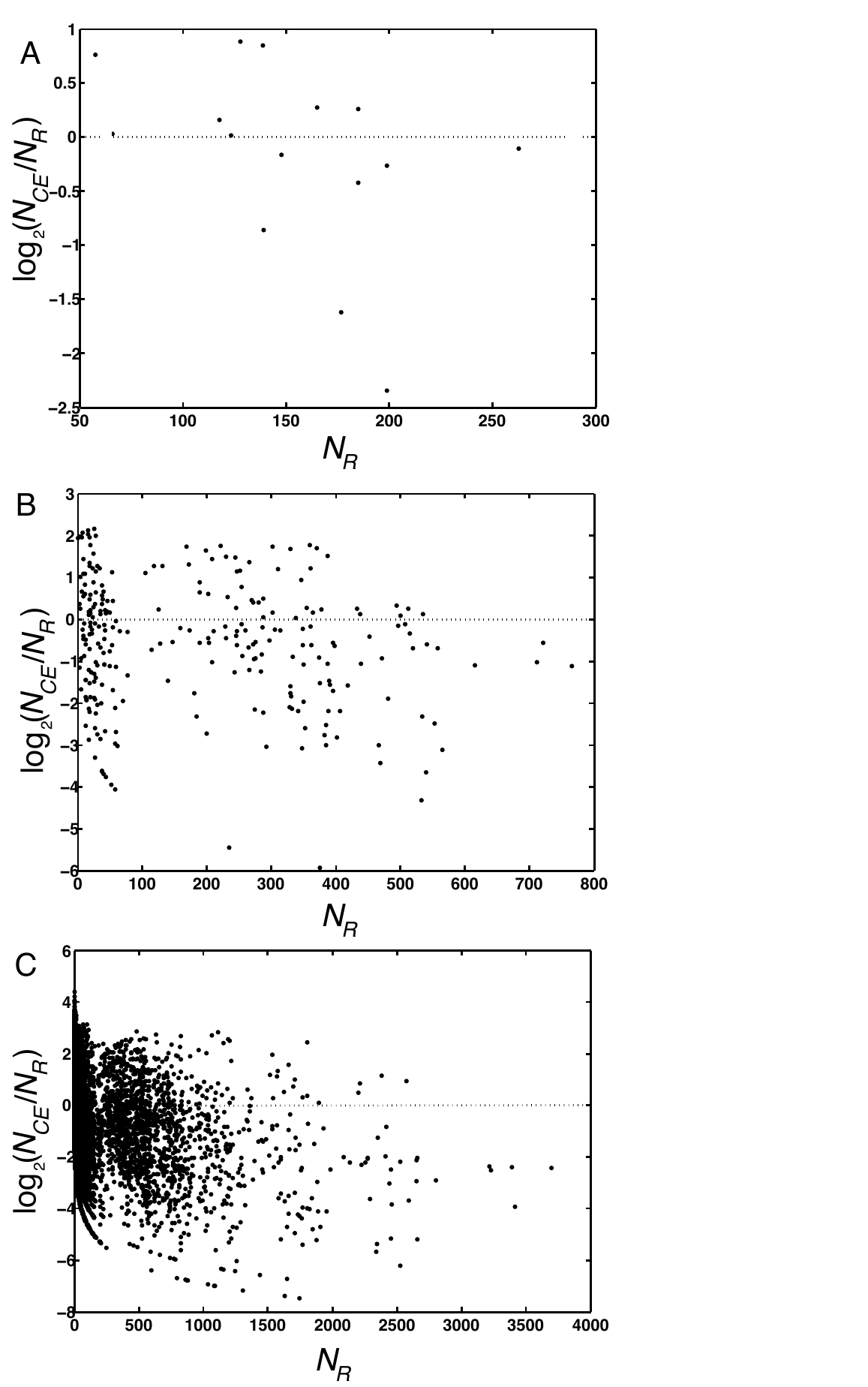} 
   \caption{{\bf Differential representation of colored motifs.} Comparison of the colored motif counts $N_{CE}$ obtained from the {\it C. elegans} neuron network and the average count from 1,000 color-randomized network, $N_R$. Points above the zero line represent the colored motifs with higher frequency in the worm's neuronal network compared to color-randomized networks (over-representation), while those below that line are suppressed. A: colored directed motifs of size 2, B: colored directed motifs with three nodes, C: colored directed motifs with 4 nodes. Logarithm is to the base 2.} 
   \label{fig:Fig2}
\end{figure}

In order to determine whether the abundance of a particular colored motif in \ce is biased, we produce random colored control networks by shuffling the color assignments in the \ce network while maintaining the relative abundance of each kind. The mean abundance $N_R$ of colored motifs of a particular type for 1,000 independent randomizations then provides the unbiased expectation for that motif, which we compare with the actual count $N_{CE}$ obtained for the colored worm brain. In Fig.~\ref{fig:Fig2}, we plot the logarithm (base 2) of the ratio $N_{CE}/N_R$ for each colored motif as a function of the random count $N_R$, to determine the extent to which the worm motifs are over- or underrepresented. Most of the motif counts in \ce are significantly different from the random control: all of the 2-node colored motif counts are significant, and all but one of the three-node motifs (one-sample two-tailed t-test, $P<0.05$). Of the 4-node motifs, only 156 of the observed 8,310 motifs are not significantly different from the control count at the 5\% level. 
We find a tendency of colored motifs in \ce to be under-represented compared to a randomly colored control, but with a significant number of motifs that are found much more often than expected by chance. (The distribution of normalized z-scores is strongly biased towards suppression, but with a long tail indicating over-expressed motifs, see Supplementary Fig. S1). This finding suggests that the majority of {\em possible} colored motifs are not useful or downright detrimental, but a handful of them are so useful that they appear between 2 and  60 times as often as in an average randomly colored network. Note that some motifs that readily appear in the random controls are completely absent in \cep: 11 colored motifs of size three and 5,460 motifs of size 4 do not appear at all, which is also significantly different from what is expected by chance: at most 5 motifs of size 3 (1.02 on average) and 3,667 motifs of size 4 (2,634 on average) were absent by chance in any of the 1,000 randomizations. 
\begin{figure}[!ht] 
   \centering
   \includegraphics[width=4.5in]{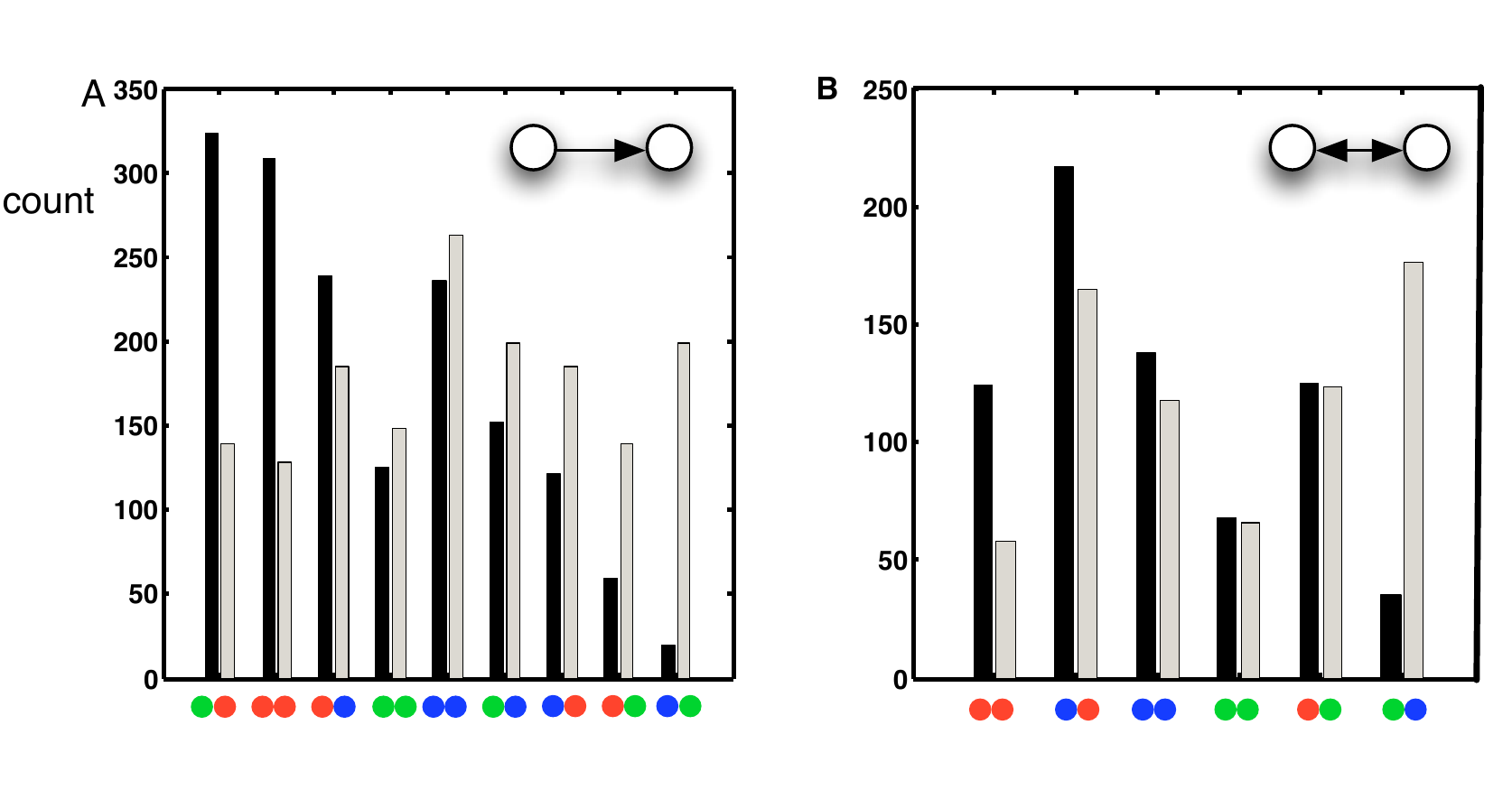} 
   \caption{{\bf Colored motif abundances.} Histogram of abundances of directed structural motifs of two neurons with particular coloration in \ce (black) compared to the average abundance in 1,000 color randomizations of the same network (grey). green: sensory neuron, red: interneuron, blue: motor neuron. A: directed pairs (the direction of information flow is left-to-right). B: undirected pairs.} 
   \label{fig:hist-2}
\end{figure}
\subsection*{Two-node motifs} In previous work that analyzed structural motifs only~\cite{SpornsKoetter2004,Reigletal2004,Songetal2005}, the undirectional two-node motif was found to be unremarkable, while the bi-directional motif was deemed over-represented~\cite{Reigletal2004,Songetal2005} with respect to an ensemble of edge-randomized networks. We can look at both of those motifs in terms of the exceptionality of their colorations. In Fig.~\ref{fig:hist-2} we show the measured counts of each of the color realizations of the directed (Fig.~\ref{fig:hist-2}A) and undirected (Fig.~\ref{fig:hist-2}B) motifs. 

These distributions show that the observed functional constraints make intuitive sense. For example, we find the motor-to-sensor-neuron motif to be significantly suppressed: we do not expect muscles to relay information to sensory neurons in a functioning worm (even though some of these connections do indeed exist). On the other hand, 
the sensor-to-inter- as well as inter-to-inter-neuron motifs appear significantly more often than expected by chance, as appropriate for information-processing motifs.

\section*{Motifs as computational building blocks}

Previous work identified the feed-forward motif as significantly over-represented~\cite{HallRussell1991,White1985,Reigletal2004,Songetal2005} in the \ce brain, as well as is gene regulatory networks~\cite{ShenOrretal2002,Miloetal2002}. We find that while many feed-forward motifs with colors are also over-represented, many others appear not to be useful. Whether a numerical over-representation (as measured, for example, by z-scores) is statistically significant must be determined carefully, by correcting for multiple hypothesis-testing (as pointed out earlier,~\cite{Reigletal2004,Songetal2005}) because it is possible that any individual motif's abundance can appear to be significantly different from the randomization control purely by chance. We have generalized the step-down min-P procedure~\cite{Reigletal2004,Songetal2005} to colored motifs (see Methods) to calculate the multiple-hypothesis-corrected P-values for the colored motifs of size 3  and 4. Using 100,000 color randomizations of the \ce network, 40 motifs of size 3 (out of a possible 273, see Table 1) have a significant P-value at the 5\% level (26 of which have a corrected $P<0.002$) and are shown in Fig.~\ref{fig:p-values}, while 505 out of the 8,310 observed motifs of size 4 have a corrected $P=0.055$ for 100,000 randomization (shown in Fig. S2). Because the number of independent hypotheses for motifs of size 4 is so large (13,770), the corrected P-values depend on the number of randomizations, and can only become significant if the number of randomizations significantly exceeds the number of hypotheses. As a consequence, the P-values for these 505 size-4 motifs will dip below the 5\% level if the number of randomizations is increased even further, and we will treat the set of 505 motifs with corrected $P=0.055$ as our set of significantly over-represented motifs of four neurons. 

\subsection*{Motifs of size three} 
\begin{figure}[!ht] 
   \centering
   \includegraphics[width=3in]{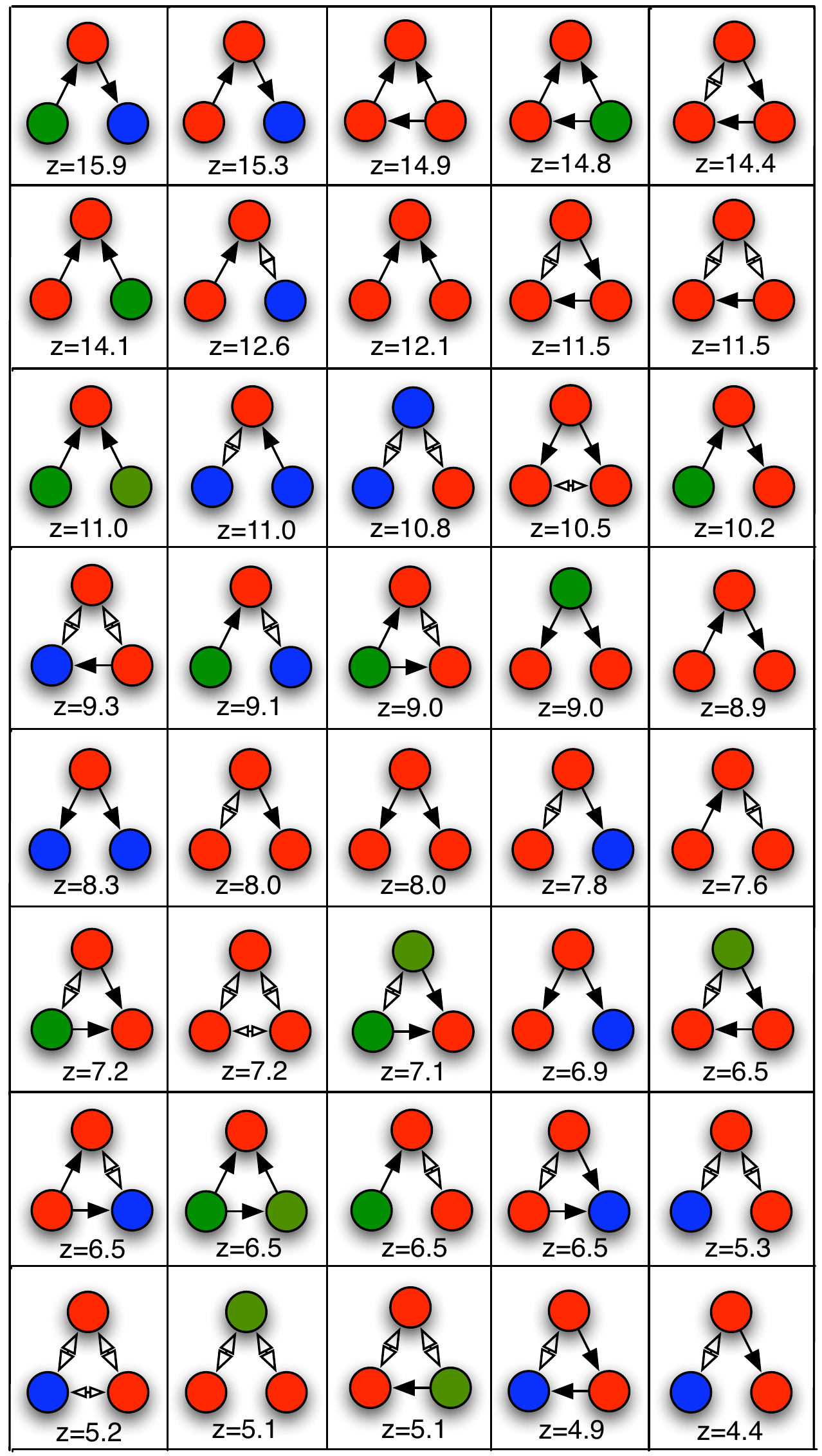} 
   \caption{{\bf Significant motifs of size 3.} Most over-represented colored motifs of size 3 in the \ce neuronal network, with their $z$-values. These motifs all have the min-P adjusted $P<0.05$.} 
   \label{fig:p-values}
\end{figure}

In Fig.~\ref{fig:p-values}, we show the 40 significantly over-represented motifs of size three, starting with 
the forward-processing motif (a relay chain) from sensor- to inter- to motor-neuron, already shown in Fig.~\ref{fig:colmotif}. That this motif is the most notable among all motifs with three neurons confirms that the overall structure of the chemical synapse network is a three-layer architecture~\cite{Varshneyetal2009}. The second-most over-used motif is also a relay-chain into the motor neuron, but from another interneuron, suggesting that this is just the 3-neuron end of a 4-neuron chain that starts with a sensor neuron. And indeed, that chain does appear among the significant motifs of size 4 (see below). The ``beginning" of that chain also appears among the significant 3-neuron motifs. The only other over-represented purely directional chain (using chemical synapses only) is the interneuron chain. 

The motif with the third-highest z-value is a feed-forward motif of three interneurons. Feed-forward motifs have different uses in computation, depending on whether the feed-forward signal is excitatory or inhibitory. Often, these motifs are used to control activation only when input is present~\cite{Miloetal2002}, or to perform ``perfect adaptation" to constant signals~\cite{Tysonetal2003}.
While feed-forward motifs have previously been identified as important in \cep, it is noteworthy that the most-used type consists of interneurons only, even though they are in the minority among neuron types. In fact, the list of motifs in Fig.~\ref{fig:p-values} is clearly dominated by interneurons, (69\% of the neurons in the list of 40 motifs, compared to only about 29\% of all neurons in the full network of 279). This imbalance suggests that the motifs represent computational building blocks that describe the information-processing task: while sensors and motors serve mainly as signal sources and sinks, interneurons work as the signal transducers.

Another highly significant feed-forward motif has the signal originating in a sensor-neuron  (see Fig.~\ref{fig:p-values}). Many of these feed-forward motifs come in alternate versions where one of the edges is an undirected gap junction, but they are never the most common. This is to be expected as there are far fewer undirected edges (514) than directed edges (2,194).
The computational purpose of a feed-forward motif using a bi-directional gap junction is not immediately obvious, but it is possible that this back connection (or many back connections) are meaningful in information-processing by providing the opportunity of feedback. Note also that the ``ring" motif where three nodes feed a signal into each other 
is absent among the significantly over-used motifs even though among three-node motifs with three edges, 40\% should be rings by chance.
While chains, splits, and merges are not significantly differentially used when analyzing uncolored motifs~\cite{Reigletal2004,Songetal2005}, some color combinations are significantly over-used as apparent from Fig.~\ref{fig:p-values}. Their purpose becomes more apparent when considering the four-node motifs.  

\subsection*{Motifs of size four}
The larger the motif, the more specific its computational function. At the same time, the number of possible motifs also increases greatly with motif-size. Of the 13,770 possible colored motifs with four nodes and directed edges, only 8,310 actually appear in the \ce network. We estimate that the number of possible colored motifs of size 5 is in the millions, preventing a significance analysis. As in the size-3 motifs, interneurons are significantly enriched within the computational motifs (68\% of neurons in size-4 significant motifs, compared to the baseline abundance of 29\% in the network as a whole).

\begin{figure}[!ht] 
   \centering
   \includegraphics[width=3.5in]{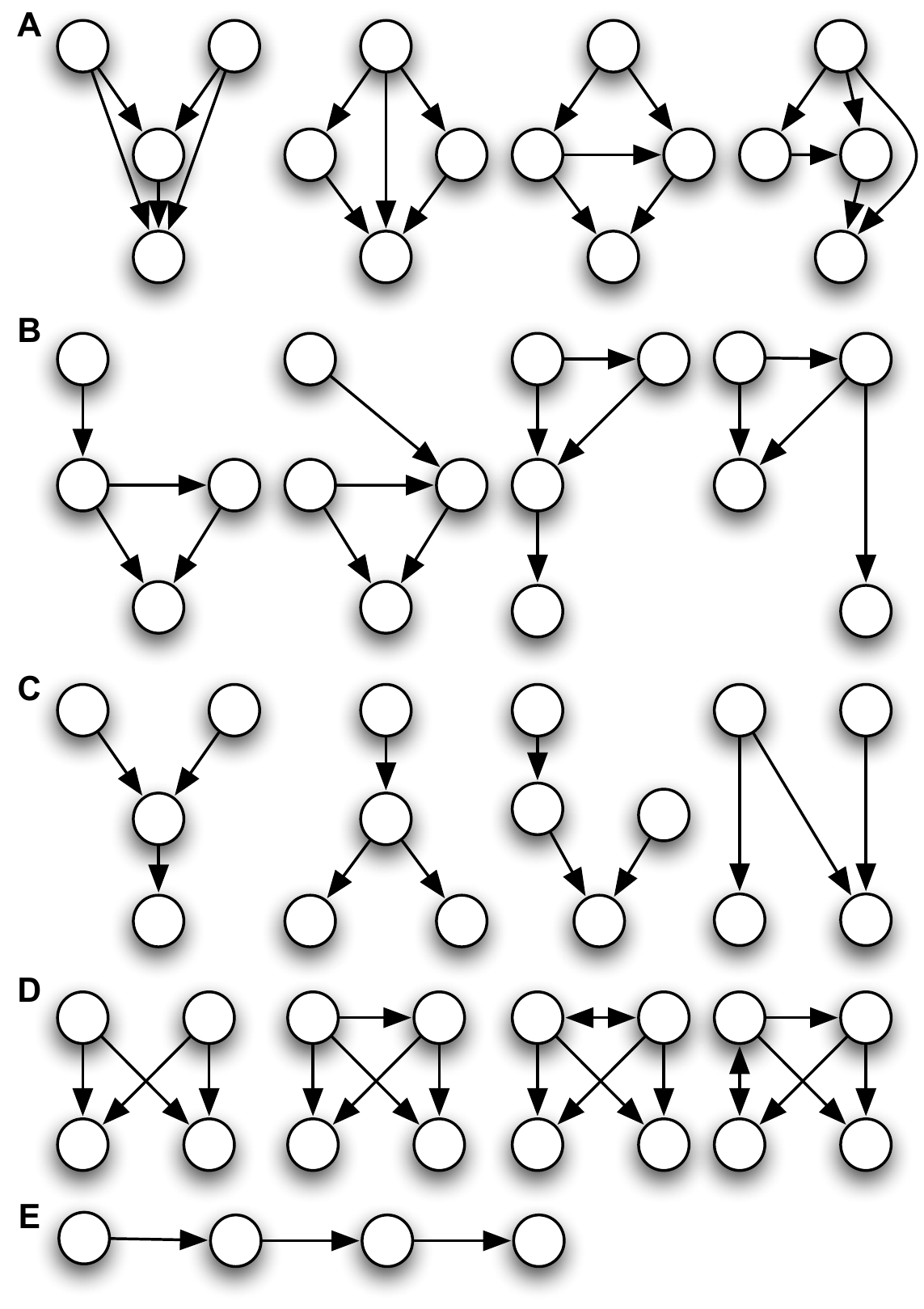} 
   \caption{{\bf Five classes of over-represented colored motifs of size 4}. A: nested feed-forward motifs, B:  feed-forward motifs with entry and exit, C: integrations and bifurcations, D: bi-fan motif with or without coupling of the inputs,  and E: linear chains.} 
   \label{fig:size-4}
\end{figure}

For four nodes with directed edges, there are 199 possible motifs that are structurally different, but many of those topologies are not prominent among the 505 colored motifs that are most significantly over-represented (shown in Supplementary Fig.~S2). Among those, we distinguish five {\em functional} classes of motifs using chemical synapses (directed edges) only, shown in Fig.~\ref{fig:size-4}. These classes cover a significant portion, but not all of the 199 possible structural motifs. (When motifs have undirected edges, they sometimes straddle two classes of motifs.) 

The most common motif-class is the nested feed-forward motif (Fig.~\ref{fig:size-4}A), of which there are several kinds. About 40\% of the most significantly over-represented motifs fall in this class. They are distinguished from bi-fan motifs (Fig.~\ref{fig:size-4}D) by the number of nodes with highest in-degree but no out-degree: bi-fans have two output nodes with an in-degree of two (see the example in Fig.~\ref{fig:loc}A) while nested feed-forwards usually have a single output node with an in-degree of two or three. Only about 5\% of the motifs among our list of 505 are bi-fans according to this definition. The second-most common group of motifs (about 25\%) are feed-forward loops with entry or exit (Fig.~\ref{fig:size-4}B), followed by the ``integration and bifurcation" motifs (Fig.~\ref{fig:size-4}C, about 20\%), and the relatively rare bi-fans, followed finally by the forward chain (5\%, Fig.~\ref{fig:size-4}E).

Functional motifs that we do not discuss (about 5\% of the motifs in the set of 505) either do not show up among the 505 prominent motifs or else are under-represented. An example is the ``nested rings" motif, an instance of which is shown in Fig.~\ref{fig:loc}B). The relative absence of ring motifs in the network could imply that feedback via the neuronal connection graph is not extensively used for computation by the worm. 

In the following, we discuss the most common colorations of the motifs in each of the classes, their possible computational function, and point out some of these motifs in a model of the \ce sub-network used for forward locomotion (Fig.~\ref{fig:loc}D), described in ~\cite{Karbowskietal2008}. 

\subsubsection*{Nested feed-forward loops}
\begin{figure}[!ht]
  \centering
   \includegraphics[width=3.5in]{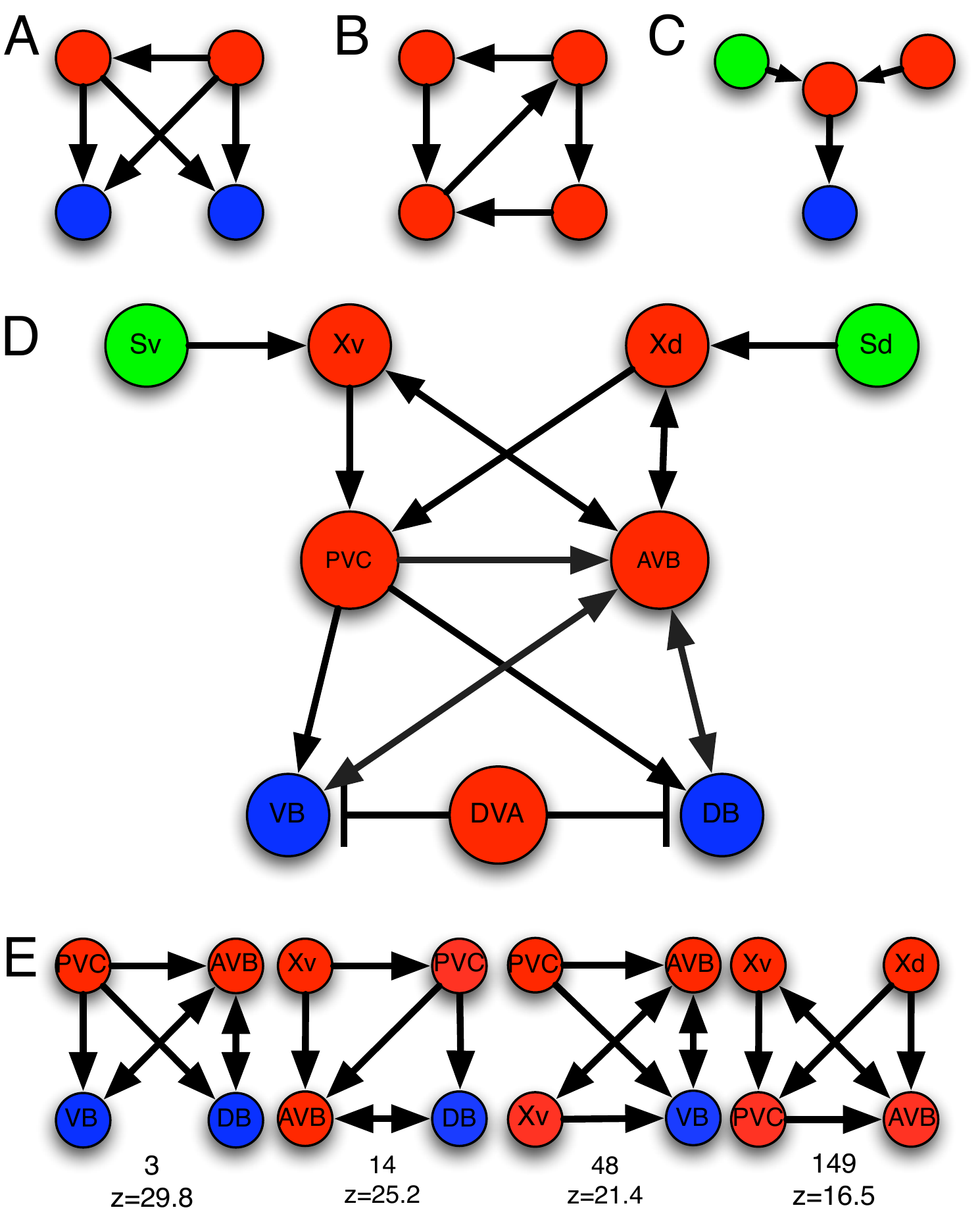} 
   \caption{{\bf Colored motifs and network context.} A: A common colored bi-fan motif. B: A nested ring motif that is uncommon in \cep. C: A signal integration motif driving a single output.  D: The core of the locomotion network of \cep, after~\protect\cite{Karbowskietal2008}. Nodes are colored according to the scheme used throughout. Arrows with single points are excitatory connections via a chemical synapse, while edges ending in a bar signal inhibition via a chemical synapse. Edges with two arrow heads denote gap junctions. E: A selection of significant four-node motifs that appear in the locomotion network shown in D. Below each motif appears the rank and z-score as in Supplementary Figure S2. Note that we included motifs that have a synaptic junction (directed link) between Xv or Xd and AVB, because both synaptic and gap junction (undirected) edges exist between those neurons.} 
   \label{fig:loc}
\end{figure}

In this class of motif, one or two inputs are fed forward through one or two relay neuron towards a single output (see examples in Fig.~\ref{fig:size-4}A). Among the top-ten colored motif types by z-value, this motif appears six times (see Supplementary Fig.~S2). We can see several motifs of this class in the reconstruction of the core network for \ce locomotion~\cite{Karbowskietal2008}, which models the undulatory behavior of the worm with a biomechanical model based on the connection structure of the \ce neuronal network. We show nine of the core network nodes and their connections in Fig.~\ref{fig:loc}D, colored according to our convention. The nodes named ``Xv" and ``Xd" are representatives of a class of interneurons (SAA) that connect in the manner shown to the ventral and dorsal head stretch receptors ``Sv" and ``Sd".  Similarly, the motor neurons labelled ``VB" and ``VD" are representatives of 18 such neurons~\cite{Wormatlas2002-2010}. The ``AVB" and ``PVC" neurons are representatives of the ``master controllers" for forward locomotion~\cite{NieburErdos1993}. The reconstruction is noteworthy because it can infer that some of the connections are inhibitory rather than excitatory. The control of PVC and AVB via the SAA neurons (Xd) in Fig.~\ref{fig:loc}D is a good example of a nested feed-forward motif, as is the control of DB via the relays PVC and AVB with Xd as the source (but note that because there are both synaptic (directed) and undirected edges between Xv and AVB, the motifs are not strictly only feed-forward). Examples of highly-represented motifs of this sort are shown in Fig.~\ref{fig:loc}E, along with their motif number and z-value as seen in Supplementary Fig.~S2. Feed-forward motifs can be nested in different manners, processing two inputs in parallel, or a single input sequentially or in a hierarchical manner (see Fig.~\ref{fig:size-4}A). All these motifs appear about equally in colorations that have sensor- or interneurons as the source, and inter- or motorneurons as the signal sink.

\subsubsection*{Feed-forward with entry or exit} 
A feed-forward loop with a signal connecting to the input, output, or relay-neuron (see sketches in Fig.~\ref{fig:size-4}B) is the 2nd most frequent structure among the most-significant colored motifs of size 4. Most commonly, the output of the feed-forward loop (the signal-neuron) is directly connected to a motor neuron, highlighting the use of the feed-forward motif in controlling locomotion (four out of the top five colored motifs in this class are of this sort). A feed-forward loop consisting only of interneurons with its output directed into a motor neuron is the motif with the two highest z-values among the 505 most-significant colored motifs. The third-most common motif in this class has the entry inter-neuron of the feed-forward loop replaced with a sensor-neuron.  Several of the motifs in this class can readily be seen in the forward locomotion network Fig.~\ref{fig:loc}D.  There are no ``ring" motifs with entry or exit among the 505 significantly over-abundant motifs of this class.  

\subsubsection*{Integrations and bifurcations}
This class of motifs (examples are depicted in Fig.~\ref{fig:size-4}C) is relatively straightforward as there is no feed-forward or feed-back of signals. Rather, signals are either distributed via bifurcations or integrated. The most common motif in this class is a sensor neuron connected to a chain of three interneurons, followed by the integration of a sensor- and an interneuron which is fed into a motor-neuron, shown in Fig.~\ref{fig:loc}C. The most common bifurcation motif is a sensor-neuron feeding an inter-neuron, whose signal is distributed over two motor-neurons. Neither of these motifs can be found within the forward locomotion network Fig~\ref{fig:loc}D or its extensions, so we assume that they are part of another important pathway for \ce behavior.

\subsubsection*{Bi-fan motif}
The bi-fan motif (Fig.~\ref{fig:size-4}D) is a well-known motif structure (see, e.g., ~\cite{Alon2007}) that regulates two independent outputs using two inputs. The computational function of any bi-fan motif depends on whether the connections are excitatory or inhibitory~\cite{Ingrametal2006}, and on whether the inputs themselves are connected. While the motif is used sparingly in \cep, some colorations are absolutely essential to the worm's behavior.  Indeed,  the bi-fan motif controlling the motor neurons VB and VD via PVC and AVB (see Fig.~\ref{fig:loc}E, rightmost motif) happens to be the third-most over-represented motif (by z-value) of all size-4 motifs (see Fig.~\ref{fig:loc}E as well as Fig.~S2). We note, however, that the version where inputs do not communicate (first type in Fig.~\ref{fig:size-4}D) is used much more rarely.

\subsubsection*{Relay chains}
Relay chains of four nodes such as depicted in Fig.~\ref{fig:size-4}E are comparatively more rare. The most over-represented such chains are the forward processing chain from a sensor- into a motorneuron via two interneurons  (which appears in Fig.~\ref{fig:loc}D) or from three interneurons into the motor neuron, followed by variations on the theme with gap junctions replacing the chemical synapses. The relative rarity of the 4-node forward chain underscores how important signal integration and feed-forward processing is for the worm.

\section*{Discussion}
Combining two sets of data that make the neuronal network of \ce one of the best understood animal control structures known, namely the connection map between neurons and the functional characterization of each neuron, allows us to gain insight into the computational building blocks of the worm brain by determining the over-representation of colored motifs, with respect to a color-randomization of a network with the connectivity unchanged. We find that while certain structural motifs have previously  been found to be significant with respect to an edge randomization of the network~\cite{Reigletal2004,Songetal2005}, many more colored motifs are highly significant. Indeed, the overall trend is the {\em suppression} of nonsensical motifs, such as signal chains where muscles feed information into sensors, or inter-neurons deliver signals to sensors. The motifs that are used significantly more often than predicted by chance, as determined by a multiple-hypothesis-corrected test, are easily identified as important elements in a signal-processing network. Sensor-neurons are almost always sources of signals (their in-degree is significantly higher than their out-degree), while motor neurons are most often the end of the signal chain. Interneurons represent a much larger fraction of nodes in computational motifs  than their overall abundance in the network would imply, suggesting that they relay the bulk of the forward-processing information. Notably absent among the common motifs are feedback loops, and relay chains longer than three neurons, underscoring the need for immediate reaction and the integration of signals. While these observations cannot take into account whether the connections between neurons are excitatory or inhibitory (as this data is not available for the majority of the connections), the comparison with the core forward-locomotion network of \ce (where this inference has been made) suggests that an analysis of colored motif utilization captures the computational processes underlying that behavior well.

In the future, we imagine that an analysis of the utilization spectrum of colored motifs can be extended to any network where nodes can be assigned tags that differentiate their biological (or social) function, but care must be taken to limit the number of functional classes as the predictive power of this approach is quickly overwhelmed when too many motifs are possible.

\section*{Methods}

\subsection*{Motif abundances and color randomization} 
The wiring diagram as well as the functional classification of neurons into sensor- inter- and motor-neurons, was obtained from~\cite{Varshneyetal2009}. Networks were encoded in terms of an adjacency matrix ${ A(i,j)}$ where ${A(i,j)=1}$ if a chemical synapse connects neuron ${ i}$ to ${ j}$, with ${ A(j,i)=0}$. Undirected edges (gap junctions) have ${ A(i,j)=A(j,i)=1}$. Colored motifs counts were obtained using our own implementation of the FANMOD algorithm~\cite{Wernicke2006,WernickeRasche2006}. We define the z-score of a \ce motif as ${z=(N_{CE}-N_R)/\sigma}$, where ${ N_{CE}}$ is the abundance for that motif in the \ce network, ${ N_R}$ is the average of the abundance distribution of that motif in color-randomized networks, and ${ \sigma}$ is the standard deviation of that distribution. 
Color randomization of the \ce colored graph is performed by repeatedly switching the colors of two randomly chosen nodes, thus preserving the color distribution and the underlying graph topology. The color switch is repeated sufficiently often to guarantee a random color distribution.

\subsection*{Multiple hypothesis testing}
We are testing the hypothesis that a colored motif in the \ce neuronal networks is significantly over-represented, compared to the same motif in a color-randomized network. Because many hypotheses are tested simultaneously, the probability of rejecting the null hypothesis for any motif by chance at least once increases with the number of hypotheses tested. To correct for this, we adapted the single-step min-P procedure for multiple-hypothesis adjustment~\cite{Dudoitetal2003,WestfallYoung1993} that was also used by \cite{Reigletal2004,Songetal2005} as follows. For each size class of motifs, let ${N_{CE}(i)}$ be the count of colored motif $i$ among the ${ M}$ possible colored motifs, and ${ N_R(p)}$ be the motif count for the same motif $i$ in the $p$th color-randomization of the \ce network (${ p=1\cdots S)}$, where ${ S}$ is the number of randomizations. Using the Heaviside (step)-function definition:
\begin{equation}  \theta(x)= \left\{\begin{array}{l} 1\ \ \ \ x>0\;,\\ 
 0\ \ \ \  x<0\;,\\
 0.5\ \ x=0\;,\end{array}\right. 
\end{equation}
we define the raw P-value for each \ce colored motif ${ i}$ as
\begin{equation}  P_{CE}(i)=\frac1S\sum_{r=1}^S \theta\left(N_R(i,r)-N_{CE}(i)\right)\;.
\end{equation}
We also define the raw P-value for any randomization $r$ of motif $i$
\begin{equation}
P(i,r)=\frac1S\sum_{p\neq r=1}^S\theta\left(N_R(i,p)-N_{R}(i,r)\right)
\end{equation}
to obtain the most significantly over-represented randomization (by chance) among the colorations $i$
\begin{equation}
P_{\min}(r)=\min_iP(r,i)\;.
\end{equation}
Finally, the single-step min-P adjusted P-value for motif ${ i}$, ${ \pi(i)}$, is obtained by comparing the raw P-value for each of the motifs $P_{CE}(i)$ to the smallest of the P-values found in the randomizations (across all motifs) as:
\begin{equation}
\pi(i)={\rm Pr}(P_{CE}(i)\leq P_{\min} (p), \ \ 1\leq p\leq S)\;.
\end{equation}

\section*{Acknowledgments}
We would like to thank Jeffrey Edlund and Paul Sternberg for discussions, and two anonymous referees for helpful comments on the manuscript. This work was supported by the National Science Foundation's Frontiers in Biological Research Grant FIBR- 0527023, and by the NSFs BEACON Center for Evolution in Action under contract No. DBI-0939454. 
\bibliography{Motifs}





\mbox{}
\pagebreak

 
\section*{Supplementary Figures}
\renewcommand{\thefigure}{S\arabic{figure}}
\setcounter{figure}{0}

\begin{figure}[!ht] 
   \centering
   \includegraphics[width=4in]{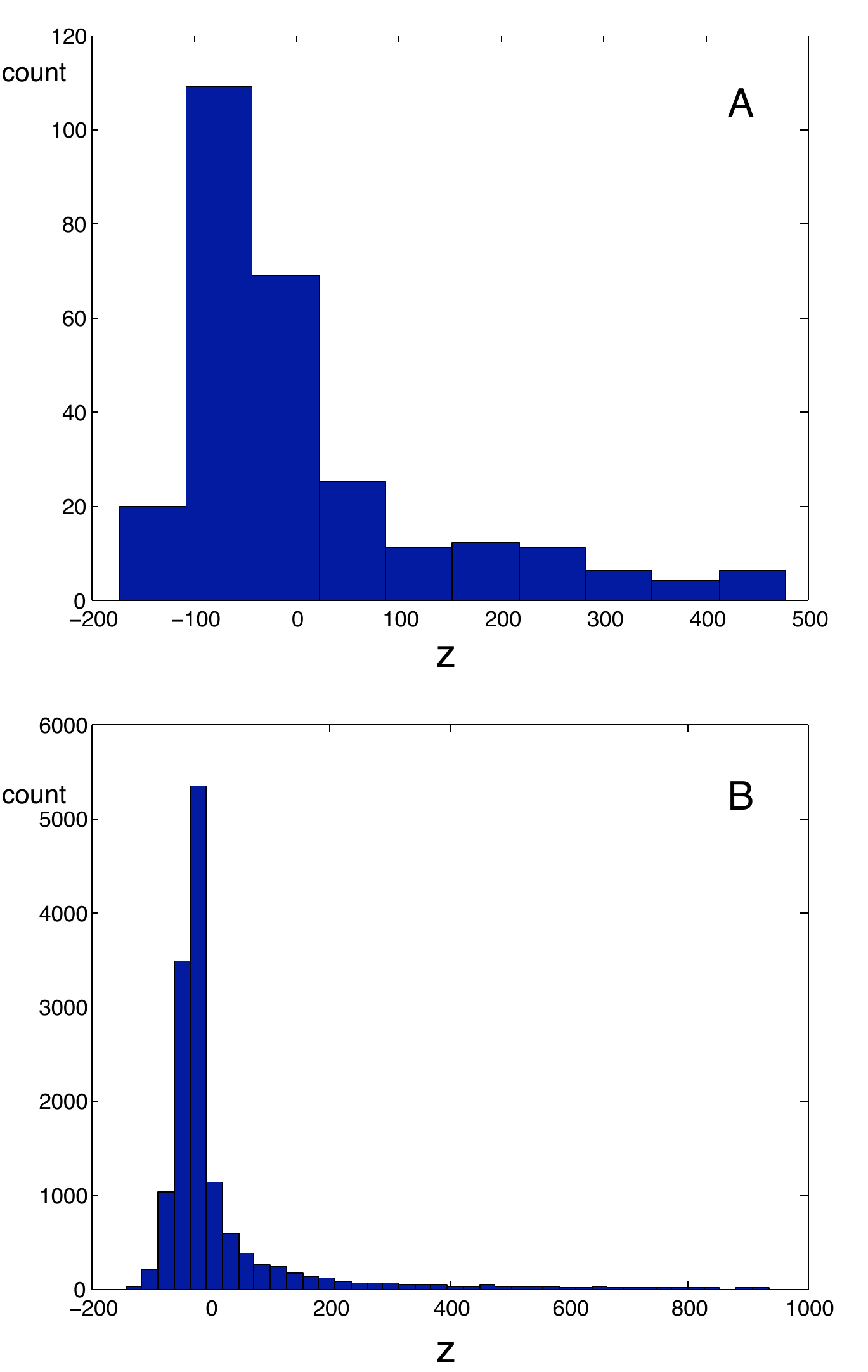} 
   \caption{Histogram of normalized $z$-scores $z=\frac{N_{CE}-N_R}{\sigma_R/\sqrt n}$ of colored motif counts, where $\sigma_R$ is the standard deviation of the count distribution with $n=1,000$ randomizations. A: motifs of size 3, B: motifs of size 4.}
   \label{fig:S1}
\end{figure}

\begin{figure}[!ht] 
   \centering
   \caption{Rank and $z$-scores (un-normalized) for the 505 motifs of size 4 with single-step min P adjusted P-value P=0.0556 for 100,000 randomizations. Each of these motifs has a $P_{CE}=0$ (see Methods), which implies each of these motifs was more abundant in the {\it C. elegans} network than in any of the 100,000 randomized networks. But because there are 5,560 entries in $P_{\rm min}$ that vanish, the adjusted P-value cannot be smaller than 0.0556. Increasing the number of randomizations leads to a smaller fraction of zeros in $P_{\rm min}$, and thus decreases the adjusted P-value of those motifs that have $P_{CE}=0$. (First 56 motifs only, remainder available upon request).}
   \label{fig:S2}
      \includegraphics[width=6in]{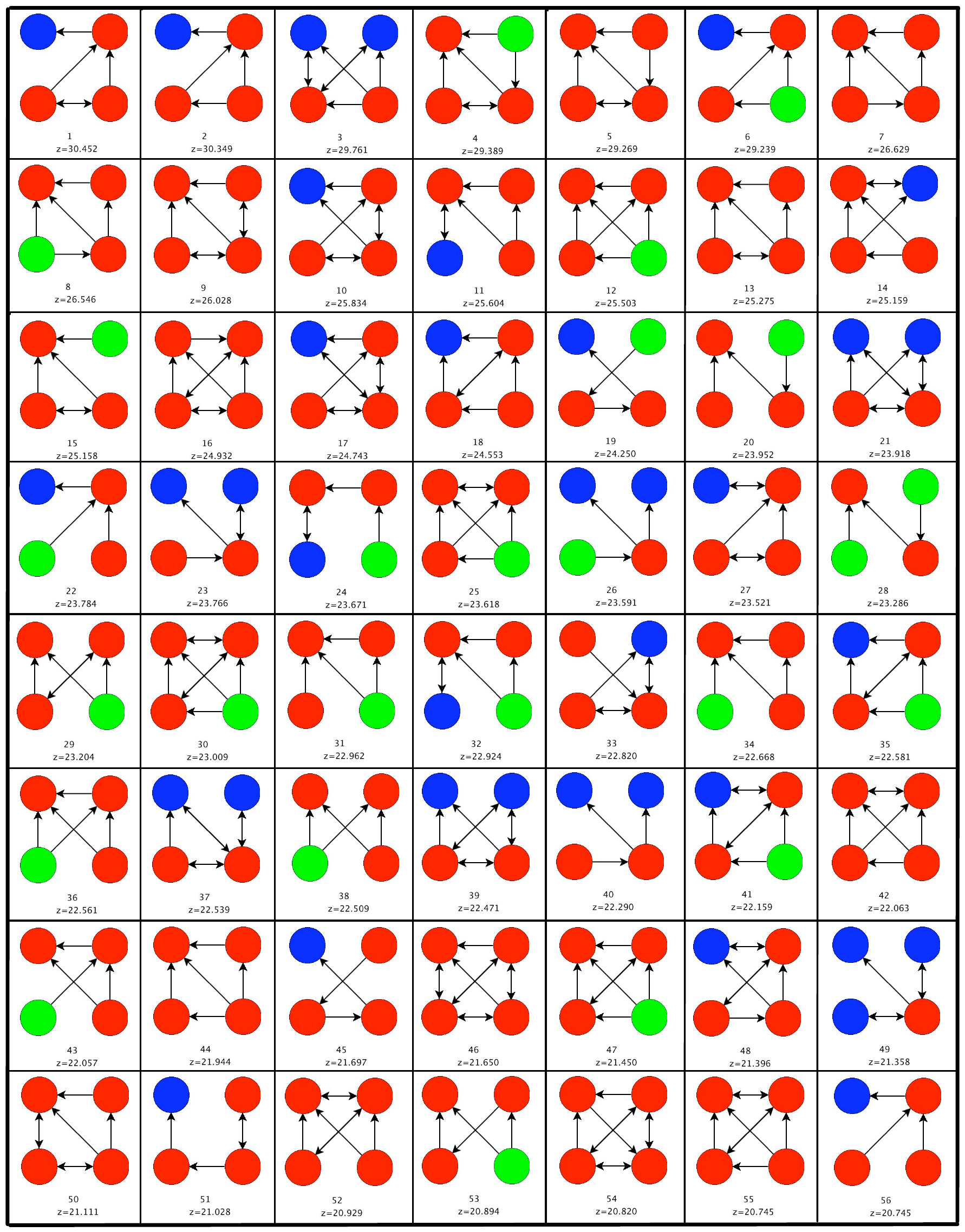} 
\end{figure}
\end{document}